\documentstyle[preprint,aps,epsfig]{revtex}
\def\be{\begin{equation}}
\def\ee{\end{equation}}
\def\ep{\epsilon}
\def\t{\tilde}

\begin{document}
\draft
\title{   Nonlinear ac response of colloidal suspension\\ with an intrinsic dispersion}
\author{J. P. Huang$^1$, L. Gao$^{1,2}$ and K. W. Yu$^1$}
\address{$^1$Department of Physics, The Chinese University of Hong Kong, \\
         Shatin, New Territories, Hong Kong}
\address{$^2$Department of Physics, Suzhou University, Suzhou 215 006, China}
\maketitle

\begin{abstract}
When a sinusoidal (ac) field is applied to a suspension containing nonlinear dielectric particles,
 the electrical response will 
generally consist of ac fields at frequencies of the higher-order 
harmonics. The situation is further complicated by an intrinsic dielectric dispersion
which often occurs due to the surface conductivity or inhomogeneous structure of the particles. 
We perform a perturbation method to investigate the effect of intrinsic dielectric dispersion
on the harmonics of local field as well as induced dipole moment.
The results showed, for weak intrinsic dispersion strength, 
the ratio of the third to first harmonics of the induced dipole moment decreases
as the frequency increases, which is qualitatively in agreement with experimental result. However,
for a strong dispersion strength, the harmonics ratio increases as the frequency increases.
Moreover, an increase in the intrinsic relaxation time may increase the strength of harmonics.

\end{abstract}
\vskip 5mm 

\section{Introduction}

When an intense electric field is applied to a suspension containing
highly polarized particled embedded in a host fluid, the particles will 
form chains due to the induced dipole moments inside the particles,
and hence complex anisotropic structures occur.
Also, the apparent viscosity of the suspension will be enhanced greatly.
These structures can have anisotropic physical properties, 
such as the effective conductivity and permittivity 
\cite{Martin-1,Martin-2}. In this case,
  optical nonlinearity enhancement was found as well \cite{Yuen}.

A convenient method of probing the nonlinear 
characteristics of the field-induced structures is to measure the 
harmonics of the nonlinear polarization under the application of 
a sinusoidal (ac) electric field \cite{Kling98}. 
Therefore it is possible to
perform a real-time monitoring of the field-induced aggregation process in such a suspension
 by measuring the optical nonlinear ac responses both parallel and perpendicular to 
the uniaxial anisotropic axis.
Recently, an anisotropic Maxwell-Garnett theory \cite{Lo} was 
applied to study the nonlinear polarization of these anisotropic structures \cite{pre-1}.
Anisotropic optical nonlinear ac response was indeed found.

When a nonlinear composite with nonlinear dielectric particles embedded
in a host medium, or with a nonlinear host medium is subjected to
a sinusoidal field, the electrical response in the composite will
in general be a superposition of many sinusoidal functions
\cite{Bergman95}. It is natural to investigate the effects of a
nonlinear characteristics on the optical nonlinear ac response in a colloidal suspension which can be 
regarded as a nonlinear composite medium \cite{Gu00,Wan}. 
The strength of the nonlinear polarization should be
reflected in the magnitude of the harmonics.
For these field-induced anisotropic structures, we found that as more and more chains are formed 
the harmonics of 
 local field and induced
dipole moment increase (decrease) for longitudinal (transverse) field case \cite{pre-1}.

On the other hand, the effect of nonlinear characteristics on the inter-particle 
force has been analyzed in a colloidal suspension of nonlinear particles 
\cite{nler-1} and further extended to a nonlinear host medium \cite{nler-2}.

Actually, the intrinsic dielectric dispersion
often occurs due to the surface conductivity or the inhomogeneous structure 
(e.g., coating effect \cite{JAP})
which often exist in colloidal suspensions. Hence it is of particular
importance to discuss the effect of an intrinsic
dielectric dispersion on the nonlinear polarization. 
In this regard, we will use a perturbation expansion method to calculate the nonlinear 
ac response of a nonlinear 
composite, i.e., harmonics of local field as well as induced dipole moment. 

\section{Nonlinear polarization and its higher harmonics}

Let us consider the system in which the
suspended particles of radius $a$ with a nonlinear dielectric constant are suspended in
a host medium of linear dielectric constant $\epsilon_2$. The
nonlinear characteristic gives rise to a field-dependent
dielectric coefficient \cite{Yu93}. In this case, the electric
displacement-electric field relation inside the spheres is given by:
\begin{eqnarray}
{\bf D}_1&=&\epsilon_1{\bf E}_1+\chi_1 |{\bf E}_1|^2{\bf E}_1
\approx \epsilon_1{\bf E}_1+\chi_1\langle |{\bf E}_1|^2\rangle{\bf E}_1\nonumber\\
         &=&(\epsilon_{1H}+\frac{\Delta\epsilon_1}{1+i\omega/\omega_0}){\bf E}_1+\chi_1\langle |{\bf E}_1|^2\rangle{\bf E}_1
  \equiv \tilde{\epsilon}_1{\bf E}_1,
\label{nl-coefficient}
\end{eqnarray}%1
with $i=\sqrt{-1}$, where $\epsilon_1$ and $\chi_1$ are the linear coefficient and the
weak nonlinear coefficient of the suspended particles, respectively.
Here $\ep_1$ is expressed as a Debye expression,
 $\ep_{1H}$ is the dielectric constant at high frequency,
$\Delta\ep_1$ is the intrinsic dispersion of the particle, 
$\omega_0=2\pi/\tau_0$ is the intrinsic characteristic frequency, with $\tau_0$ being
the intrinsic relaxation time, and $\omega$ is the angular frequency of the external field.
 In Eq.(\ref{nl-coefficient}), we have 
adopted an approximation: the local field inside the
particles is assumed to be uniform. This assumption is called
the decoupling approximation \cite{Yu96}. It has been shown that
such an approximation yields a lower bound for the accurate result
for the local field \cite{Yu96}. We further assumed that $\chi_1$
is independent of frequency, which is a
valid assumption for low-frequency processes in colloidal suspensions. As a
result, the induced dipole moment under an applied field 
${\bf E}(t)=E(t)\hat{\bf z}$ along $z-$axis, with $t$ being time, is given by \cite{pre-1}:
\begin{equation}
\widetilde{{\bf p}}=\tilde{\epsilon}_e a^3 \tilde{b} {\bf E}(t),
\end{equation}%2
where $\t{\ep}_e$ is the effective dielectric constant
of the suspension, and $\tilde{b}$ is the field-dependent dipole factor:
\begin{equation}
\tilde{b}={\tilde{\epsilon_1}-\epsilon_2 \over
\tilde{\epsilon_1}+2\epsilon_2} ={\epsilon_1+\chi_1\langle
|{\bf E}_1|^2 \rangle-\epsilon_2 \over \epsilon_1+\chi_1\langle |{\bf E}_1|^2
\rangle+2\epsilon_2}.
\end{equation}%3

In order to obtain the effective dielectric constant $\t{\ep}_e$, we 
invoke the Maxwell-Garnett approximation (MGA) for anisotropic 
composites \cite{Lo,pre-1}.
For the longitudinal field case when the ac field is applied along
the uniaxial anisotropic axis, the MGA has the form
 \be 
\frac{\t{\ep}_e-\ep_2}{\beta_L\t{\ep}_e+(3-\beta_L)\ep_2}=f
\frac{\t{\ep}_1-\ep_2}{\t{\ep}_1+2\ep_2}.
\label{MGA}
 \ee %4
with $f$ being the volume fraction of particles,
whereas for an transverse field case when the ac field is applied 
perpendicular to the uniaxial anisotropic axis, the MGA expression is obtained by replacing
$\beta_L$ with $\beta_T=(3-\beta_L)/2$ \cite{Lo,pre-1}. 
Here $\beta_{L}$ and $\beta_{T}$ denote the local field factors parallel
and perpendicular to the uniaxial anisotropic axis. 
These local field factors are defined as the ratio of the local field
in the particles to the Lorentz cavity field \cite{Lo}.
For isotropic composites, $\beta_L=\beta_T=1$, while both $\beta_L$ and 
$\beta_T$ will deviate from unity for an anisotropic distribution of
particles in composites.
These $\beta$ factors have been evaluated in a tetragonal lattice of
dipole moments \cite{Lo} and in various field-structured composites 
\cite{Martin-3}.

We resort to the spectral representation approach \cite{Bergman}
to represent the local field inside the particle \cite{Yuen}:
\be
\langle |{\bf E}_1|^2 \rangle=\frac{{\bf E}^2(t)}{f}\int_0^1\frac{|s|^2\mu(x)}{|s-x|^2}{\rm d}x,
\label{EE-1}
\ee
with the material parameter $s=(1-\t{\ep}_1/\ep_2)^{-1}$ as well as
 the spectral density function $\mu(x)=f\delta[x-(1-f\beta)/3]$,
 which satisfies the sum rule $\int_0^1\mu(x){\rm d}x=f$. 
Hence Eq.(\ref{EE-1}) has the form
\be
\langle |{\bf E}_1|^2 \rangle=\frac{9|\ep_2|^2}{|\t{\ep}_1(1-f\beta)+\ep_2(2+f\beta)|^2}{\bf E}^2(t).
\label{average1}
\ee
Actually, this is a self-consistent 
equation for $\langle|{\bf E}_1|^2\rangle$, which can be solved at least numerically \cite{pre-1}. 
In what follows, we will use a 
perturbation method instead to obtain analytic expressions for harmonics of the local electric field and the induced
dipole moment.

On the other hand, due to the uniform local field in each particle, Eq.(\ref{average1})
 can also be directly derived from \cite{Yu93,ssc}:
\be
\langle |{\bf E}_1|^2\rangle=\frac{1}{f}{\bf E}^2(t)\left |\frac{\partial\t{\ep}_e}{\partial\t{\ep}_1}\right |.
\ee

If we apply a sinusoidal electric field, i.e. ${\bf E}(t)={\bf E}\sin(\omega t)$, 
the induced dipole moment $\t{p}$ will depend on time sinusoidally, too.
By virtue of the inversion symmetry, $\t{p}$ is a superposition of 
odd-order harmonics such that 
\be
\t{p}=p_{\omega}\sin \omega t+p_{3\omega}\sin 3\omega t+ \cdots.
\ee
Also, the local electric field contains similar harmonics
\be
\sqrt{\langle |{\bf E}_1|^2\rangle }=E_{\omega}\sin \omega t+E_{3\omega}\sin 3\omega t+ \cdots.
\ee
These harmonic coefficients can be extracted from the time dependence
of the solution of $\t{p}$ and $E_1(t)$.

\section{Analytic Solutions}
In this section, we will apply the perturbation
expansion \cite{Gu00,Gu92} to extract the harmonics of 
the local electric field and the induced dipole moment.
It is known that the perturbation expansion method is applicable to weak 
nonlinearity only, limited by the convergence of the series expansion.
 
We expand $\t{p}$ and $\sqrt{\chi_1\langle |{\bf E}_1|^2\rangle }$ into a Taylor expansion,
taking $\chi_1\langle |{\bf E}_1|^2 \rangle$ as a perturbative quantity:
\be
\t{p}=\sum_{n=0}^{\infty}\frac{a^3E(t)}{n!}\frac{\partial^n}{\partial\t{\ep}_1^n}
\left [\ep_2\t{b}\frac{\t{\ep}_1(1+3f-f\beta)+\ep_2(2-3f+f\beta)}
{\t{\ep}_1(1-f\beta)+\ep_2(2+f\beta)} \right ]_{\t{\ep}_1=\ep_1}(\chi_1\langle |{\bf E}_1|^2\rangle )^n ,
\ee
\be
\sqrt{\chi_1\langle |{\bf E}_1|^2\rangle }=
\sum_{m=0}^{\infty}\frac{1}{m!}\frac{\partial^m}{\partial\t{\ep}_1^m}
\left [\frac{3|\ep_2|\sqrt{\chi_1{\bf E}^2(t)}}{|(1-f\beta)\t{\ep}_1+(2+f\beta)\ep_2|}\right ]_{\t{\ep}_1=\ep_1}(\chi_1\langle |{\bf E}_1|^2\rangle )^m.
\ee
In view of weak nonlinearity, we can rewrite Eqs.(10) and 
(11), keeping the lowest orders of $\chi_1\langle |{\bf E}_1|^2 \rangle$ and $\chi_1 {\bf E}^2(t)$:
\begin{eqnarray}
\t{p}&\approx &h_1E(t)+h_3\chi_1 E^3(t),\nonumber\\
\sqrt{\chi_1\langle |{\bf E}_1^2|\rangle }&\approx &j_1\sqrt{\chi_1{\bf E}^2(t)}+j_3\left (\sqrt{\chi_1{\bf E}^2(t)}\right )^3, 
\end{eqnarray}
where
\begin{eqnarray}
h_1&=&a^3\ep_2\frac{\ep_1-\ep_2}{\ep_1+2\ep_2}\frac{\rho+3f(\ep_1-\ep_2)}{\rho},\nonumber\\
h_3&=&\frac{27a^3|\ep_2|^2\ep_2^2}{\rho|\rho|^2(\ep_1+2\ep_2)}\left [\frac{3f(\ep_1-\ep_2)}{\rho}+
\frac{\rho+3f(\ep_1-\ep_2)}{\ep_1+2\ep_2}\right ],\nonumber\\
j_1&=&\frac{3|\ep_2|}{\rho},\ \ 
j_3=\frac{27}{2}|\ep_2|^3\frac{f\beta-1}{\rho|\rho|}(\frac{1}{|\rho|^2}+\frac{1}{(\rho^{2})^*}),\nonumber
\end{eqnarray}
with $\rho=(1-f\beta)\ep_1+(2+f\beta)\ep_2$.

In the case of a sinusoidal field, we can expand $E^3(t)$ in terms of 
the first and the third harmonics.
The comparison with Eq.(8) yields the harmonics of the induced dipole moment:
\be
p_{\omega} = h_1 E+\frac{3}{4}h_3\chi_1 E^3, \ \ \
p_{3\omega} = -\frac{1}{4}h_3 \chi_1 E^3.
\label{p3w}
\ee
Similarly, we find the harmonics of the local electric field
\be
\chi_1^{1/2}E_{\omega} = j_1\chi_1^{1/2}E+\frac{3}{4}j_3(\chi_1^{1/2}E)^3,
\ \ \
\chi_1^{1/2}E_{3\omega} = -\frac{1}{4}j_3(\chi_1^{1/2}E)^3.
\label{e3w}
\ee

In the above analysis, we have used the identity 
$\sin^3\omega t=(3/4)\sin \omega t-(1/4)\sin 3\omega t$ 
to obtain the first and the third harmonics. 
Similar analysis can be used to extract the higher-order harmonics, 
by retaining more terms in the series expansion. For instance, to obtain fifth harmonics,
we may take into account the identity
$\sin^5\omega t=(5/8)\sin\omega t-(5/16)\sin 3\omega t+(1/16)\sin 5\omega t$, and then obtain the
harmonics of induced dipole moment and local electric field, respectively:
\begin{eqnarray}
p_{\omega}&=&h_1E+\frac{3}{4}h_3\chi_1E^3+\frac{5}{8}h_5\chi_1^2E^5,\nonumber\\
p_{3\omega}&=&-\frac{1}{4}h_3\chi_1E^3-\frac{5}{16}h_5\chi_1^2E^5,\nonumber\\
p_{5\omega}&=&\frac{1}{16}h_5\chi_1^2E^5,
\end{eqnarray}
where
\begin{eqnarray}
h_5&=&\frac{81}{2}\frac{a^3|\ep_2|^4}{|\rho|^4}\{ \frac{9f\ep_2^2}{\rho^2(\ep_1+2\ep_2)}
[\frac{-2(1-f\beta)(\ep_1-\ep_2)}{\rho}+\frac{3\ep_2}{\ep_1+2\ep_2}]+
\frac{3\ep_2^2}{\rho(\ep_1+2\ep_2)^2}\nonumber\\
& &[\frac{(1+3f-f\beta)\rho-(1-f\beta)(\rho+3f\ep_1-3f\ep_2)}{\rho}-
\frac{2\rho+6f(\ep_1-\ep_2)}{\ep_1+2\ep_2}] \},\nonumber
\end{eqnarray}
and
\begin{eqnarray}
\chi_1^{1/2}E_{\omega}&=&j_1\chi_1^{1/2}E+\frac{3}{4}j_3(\chi_1^{1/2}E)^3+
\frac{5}{8}j_5(\chi_1^{1/2}E)^5,\nonumber\\
\chi_1^{1/2}E_{3\omega}&=&-\frac{1}{4}j_3(\chi_1^{1/2}E)^3-\frac{5}{16}j_5(\chi_1^{1/2}E)^5,\nonumber\\
\chi_1^{1/2}E_{5\omega}&=&\frac{1}{16}j_5(\chi_1^{1/2}E)^5,\nonumber
\end{eqnarray}
where
$$
j_5=\frac{243}{4}\frac{(1-f\beta)^2|\ep_2|^5}{|\rho|^4}(\frac{3}{2}\frac{1}{\rho^2|\rho|}+
\frac{1}{|\rho|^3}+\frac{3}{2}\frac{\rho^2}{|\rho|^5}).
$$
Because of the weak nonlinearity under consideration, the effect of high order harmonics  
is quite small. Thus, without loss of generality, we only consider the 
Taylor expansion up to second term. In this regard, 
we will use Eqs.(\ref{p3w}) and (\ref{e3w}) to investigate the   nonlinear ac response 
in the numerical calculations in next section.

\section{Numerical Results}

We are now in a position to perform numerical calculations to investigate
the dispersion effect on the harmonics of local electric field as well as 
induced dipole moment. We let $f=0.15$, $\ep_{1H}=6\epsilon_0$, 
$\ep_2=75\epsilon_0$, where $\epsilon_0$ is the dielectric constant of 
free space. Additionally, set ${\bf p}_0=\ep_e a^3(\ep_1-\ep_2)/(\ep_1+2\ep_2){\bf E}$
to normalize the induced dipole moment, where $\ep_e$ is given by the anisotropic MGA (Eq.(\ref{MGA}))
without nonlinear characteristic. Obviously, the expressions for 
the normalized local electric field and the normalized 
induced dipole moment are functions of a single variable $\chi_1^{1/2}E$ \cite{Wan}. 
In view of the weak nonlinearity, we set $\chi_1^{1/2}E=0.9\sqrt{\epsilon_0}$ for our calculation.
  
In Fig.1(a)$\sim$(c), we investigate the harmonics of local field and induced dipole moment versus frequency
for different intrinsic dispersion strengths $\Delta\ep_1$.
It is obvious that increasing the dispersion strength reduces
the harmonics of both the local field and the induced dipole moment. 
We find that, for increasing frequency, the harmonics of local field always increases. 
However, the harmonics of induced dipole moment exhibits a more complicated behavior.
For small $\Delta\ep_1$, the ratio of the third to first harmonics
of induced dipole moment decreases as the frequency increases,
 which is qualitatively in good agreement with
the experimental data \cite{Kling98}. However, for large $\Delta\ep_1$, its harmonic ratio
 also increases
as the frequency increases. We expect this behavior can be demonstrated by experiment in the future.

In Fig.2, we investigate the harmonics of local field and induced dipole moment versus frequency
for different intrinsic relaxation times $\tau_0$. 
A longitudinal field case ($\beta_L=2$) is discussed in Fig.2(a), while a transverse field case
($\beta_T=0.5$) in Fig.2(b). It is shown that, in either case,
increasing relaxation time $\tau_0$ may enhance the harmonics of both local field and
induced dipole moment. In addition, the transverse field case may predict stronger
harmonics than the longitudinal field case within high-frequency region,
which is actually valid for different dispersion strengths as well (no figures shown herein).

On the other hand, Klingenberg claimed that increasing external electric field leads 
to increasing harmonics of electric 
current \cite{Kling98}. In this sense, we have also calculated
 the effect of external electric field on the   nonlinear ac response of
the induced dipole moment (no figures shown herein). 
 Obviously, our calculations  qualitatively demonstrate
the experimental results \cite{Kling98} as well.

\section{Discussion and conclusion}

Here a few comments on our results are in order. In the present study, 
we have examined the case of nonlinear particles suspending in a linear 
host. Our considerations may be extended to a nonlinear host medium 
\cite{nler-2}. 

We have studied the effect of an intrinsic dispersion on 
the   nonlinear ac response of a colloidal suspension.
Actually, our formalism
can be readily generalized to discuss two or more intrinsic dielectric dispersions of particles 
(or host fluid).
 In view of the fact that the intrinsic 
dielectric dispersion
 often occurs due to the surface conductivity or the inhomogeneous structure (e.g., coating effect \cite{JAP}), 
 we believe such effect also plays an important role in the dielectric behavior of
 biological cells, such as the electrorotation, dielectrophoresis and electro-orientation.

The perturbation method is used in the present work, but 
we believe similar results may also be predicted by a self-consistent theory \cite{pre-1}.

In summary, a perturbation method has been employed to
compute the local electric field and the induced dipole moment for suspensions in which the
suspended particles have a nonlinear characteristic and an intrinsic dispersion, in an
attempt to investigate the frequency effect on the   nonlinear ac response. The results showed, 
for weak intrinsic dispersion strength, 
the ratio of the third to first harmonics of the induced dipole moment decreases
as the frequency increases, which is qualitatively in agreement with experimental result. However,
for a strong dispersion strength, the harmonics ratio increases as the frequency increases.
Moreover, an increase in the intrinsic relaxation time may increase the strength of harmonics.

\section*{Acknowledgments}
This work was supported by the Research Grants Council of the Hong Kong SAR
Government under project numbers CUHK 4284/00P and CUHK 4245/01P. K.W.Y. thanks Prof. A. R. Day
for his interest in our work and his suggestion to examine the effects of inhomogeneities
on the dielectric response of biological cell suspensions.

\begin{figure}[h]
\caption{Harmonics of the local electric field and the induced dipole moment versus frequency for 
$\tau_0=10^{-6.5}s$ and $\beta_L=2$, for different intrinsic dispersion strength. 
(a) $\Delta\epsilon_1=50\epsilon_0$;
(b) $\Delta\epsilon_1=100\epsilon_0$;
(c) $\Delta\epsilon_1=150\epsilon_0$.}
\end{figure}

\begin{figure}[h]
\caption{Harmonics of the local electric field and the induced dipole moment versus frequency for 
$\Delta\epsilon_1=50\epsilon_0$, for different intrinsic relaxation time $\tau_0$. 
(a) $\beta_L=2$; 
(b) $\beta_T=0.5$.}
\end{figure}

\newpage
\centerline{\epsfig{file=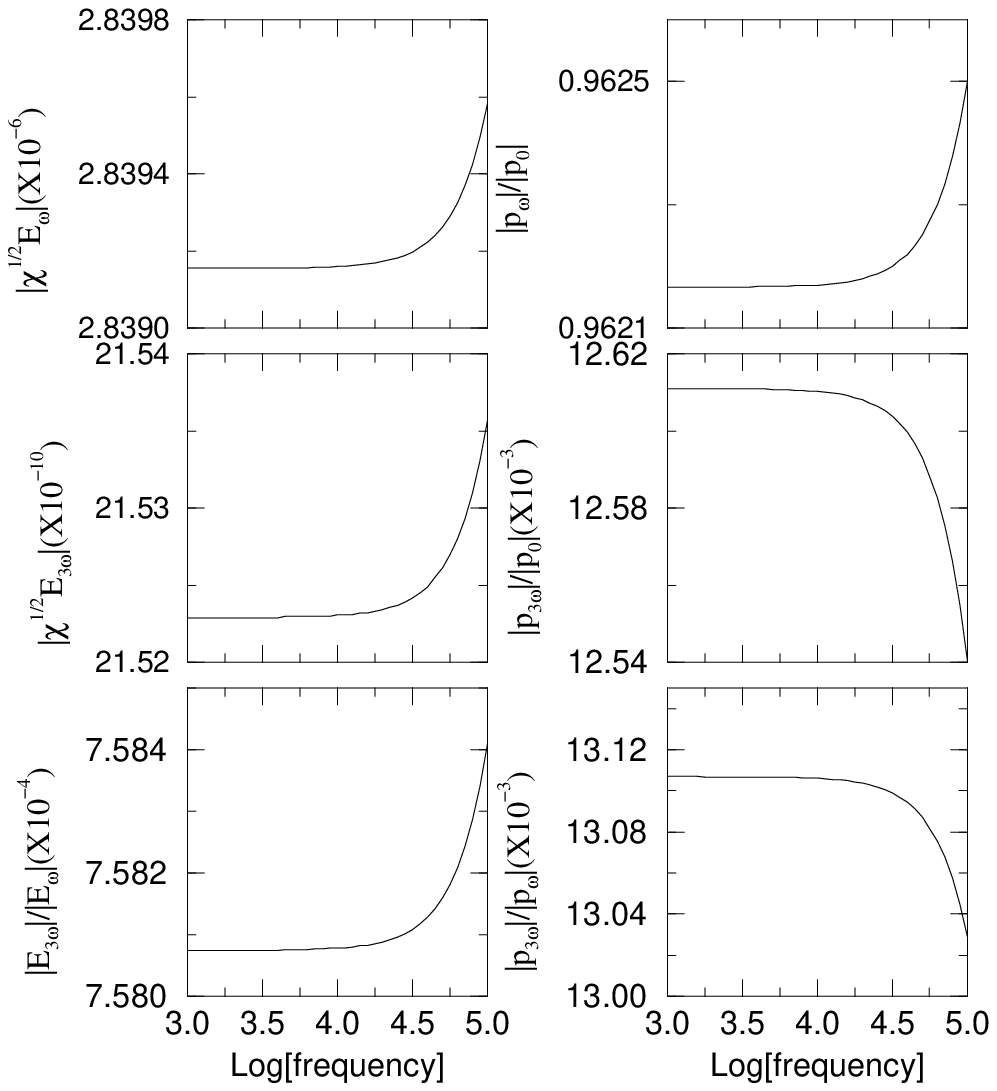,width=\linewidth}}
\centerline{Fig.1 (a)}

\newpage
\centerline{\epsfig{file=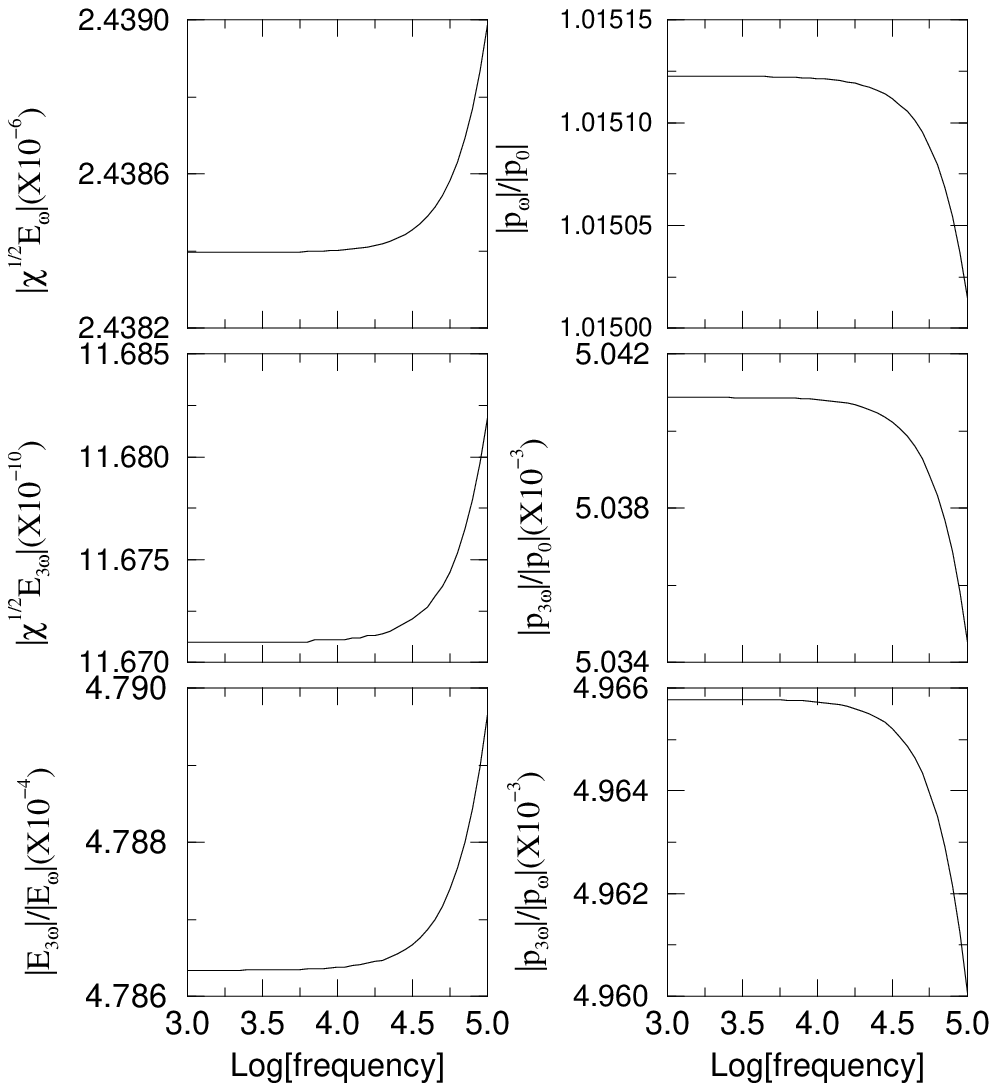,width=\linewidth}}
\centerline{Fig.1 (b)}

\newpage
\centerline{\epsfig{file=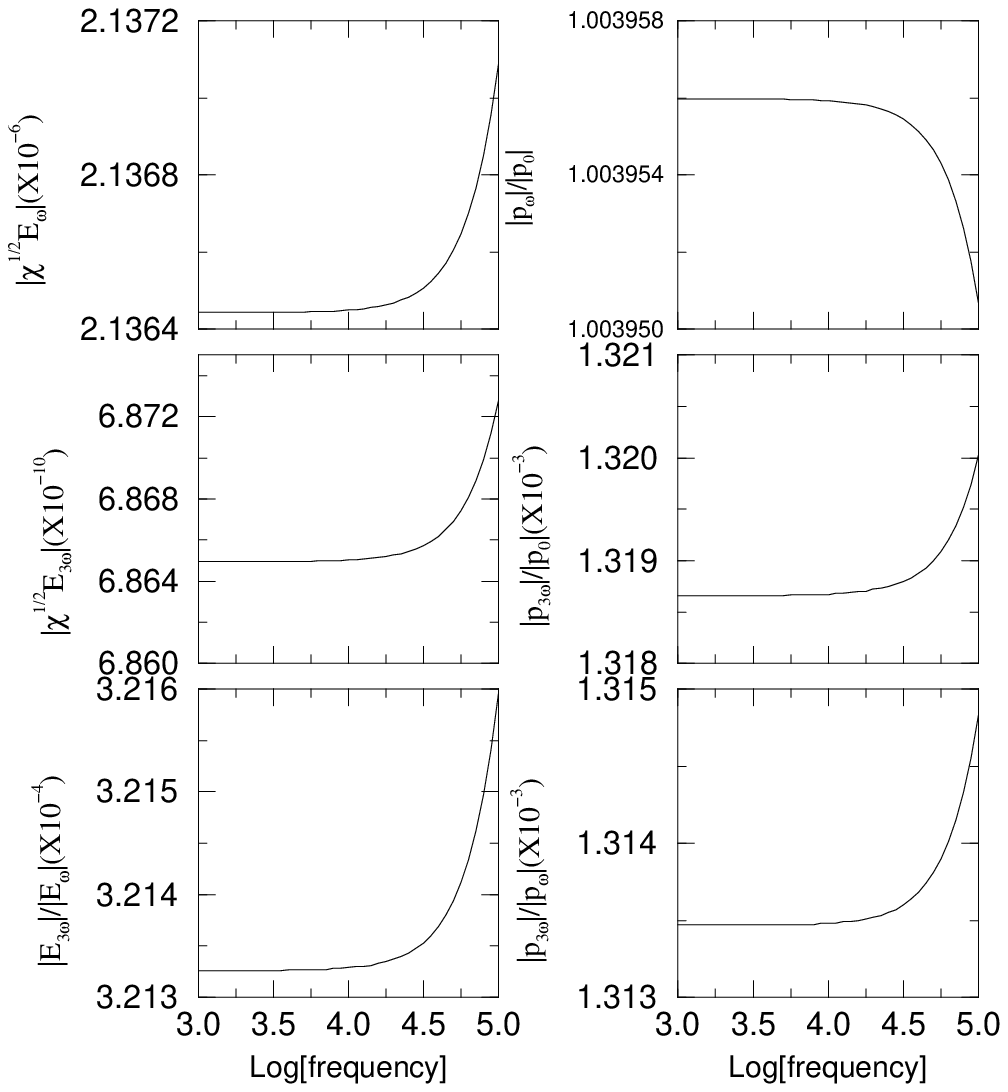,width=\linewidth}}
\centerline{Fig.1 (c)}

\newpage
\centerline{\epsfig{file=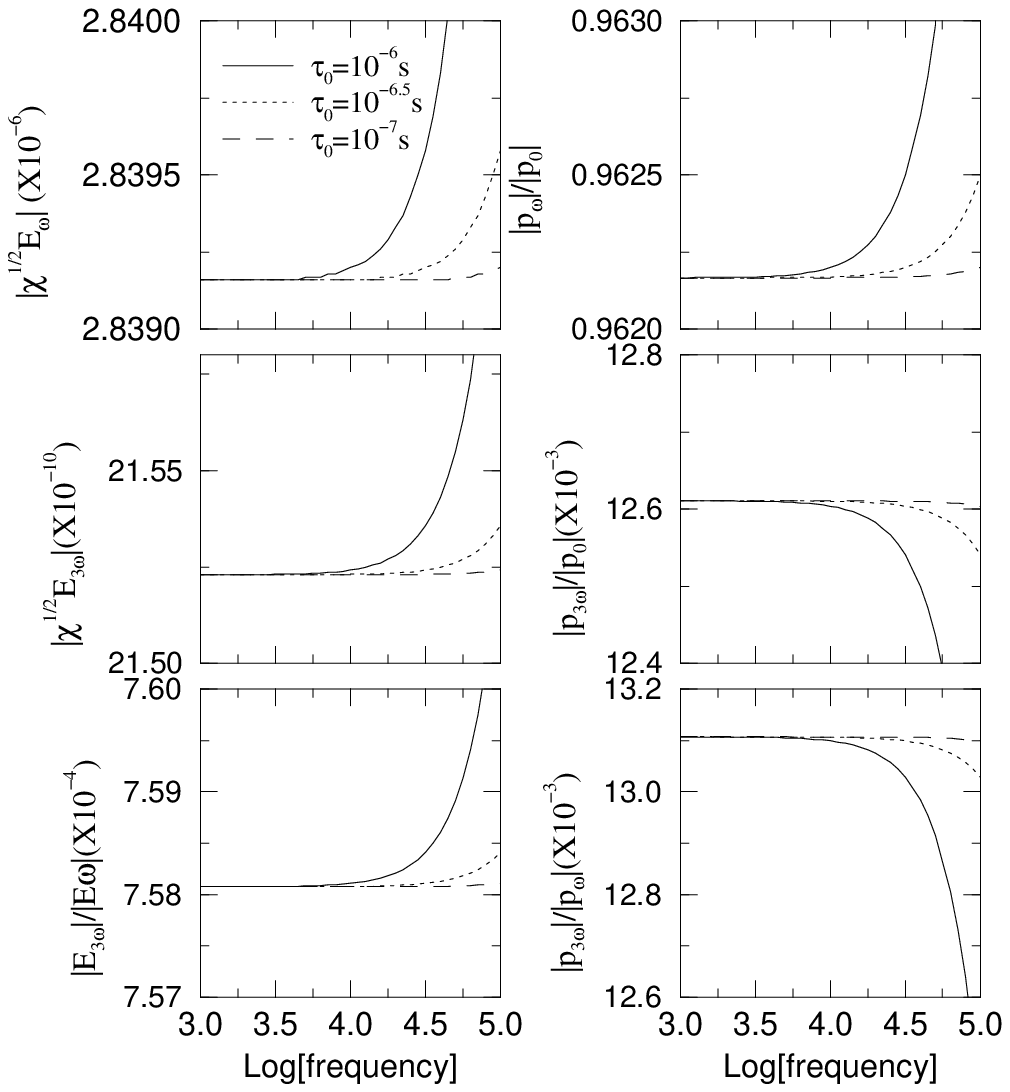,width=\linewidth}}
\centerline{Fig.2 (a)}

\newpage
\centerline{\epsfig{file=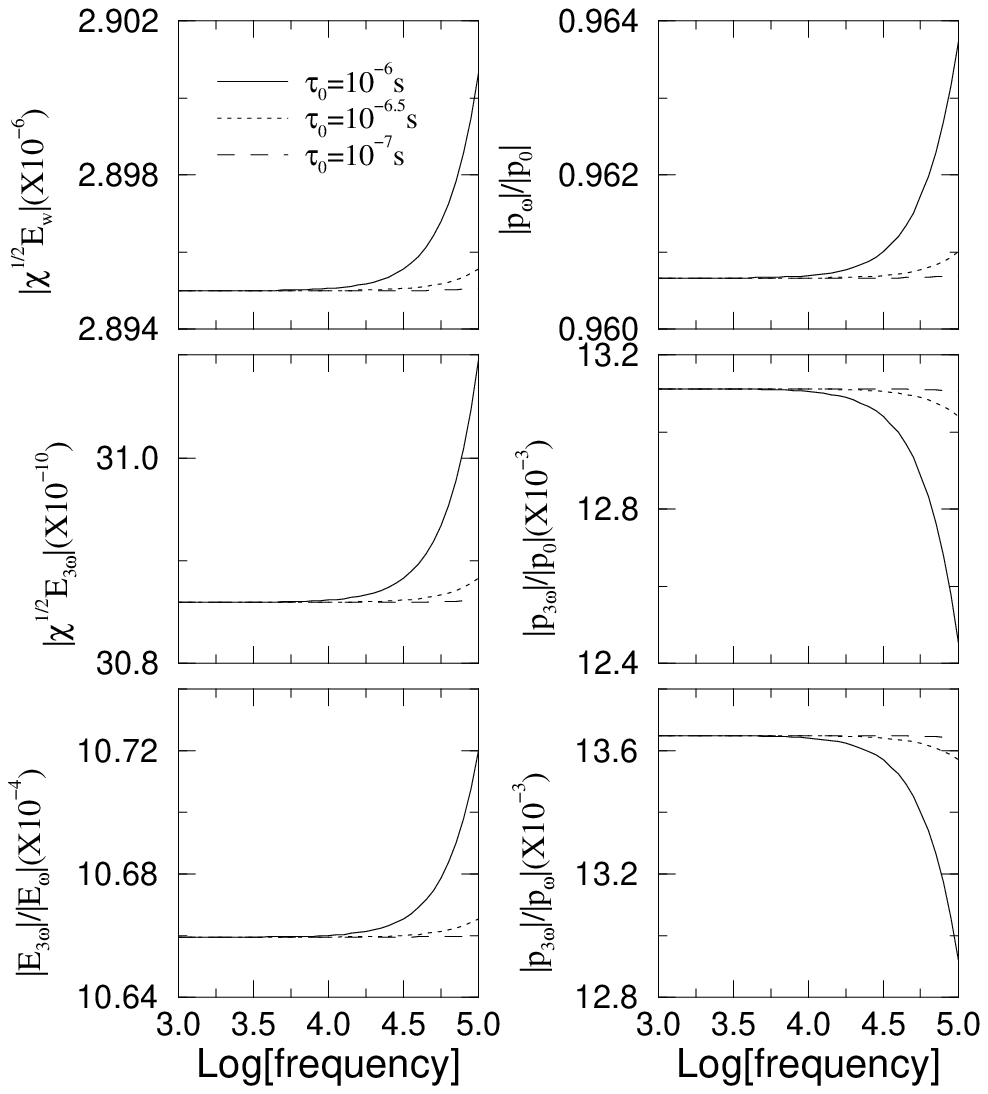,width=\linewidth}}
\centerline{Fig.2 (b)}

\end{document}